\documentclass[prb,
                    preprint
                    ]{revtex4}
\usepackage{epsfig}

\def\e{\begin{equation}}
\def\f{\end{equation}}
\def\l#1{\label{eq:#1}}
\def\r#1{(\ref{eq:#1})}

\begin{document}

\title{Near-field enhancement and sub-wavelength imaging in the optical region using a pair of two-dimensional arrays of metal nanospheres}

\author{P. Alitalo$^1$}

\author{C. Simovski$^{1,2}$}

\author{A. Viitanen$^3$}

\author{S. Tretyakov$^1$}

\affiliation{$^1$Radio Laboratory / SMARAD Center of Excellence, Helsinki University of Technology, P.O. Box 3000, FI-02015 TKK, Finland\\
$^2$Physics Department, State University of Information Technologies, Mechanics and Optics, St. Petersburg, Russia\\
$^3$Electromagnetics Laboratory, Helsinki University of Technology, Finland\\
{\rm E-mails: pekka.alitalo@tkk.fi, simovsky@phoi.ifmo.ru,
ari.viitanen@tkk.fi, sergei.tretyakov@tkk.fi}}

\date{\today}

\begin{abstract}
Near-field enhancement and sub-wavelength imaging properties of a
system comprising a coupled pair of two-dimensional arrays of
resonant nanospheres are studied. The concept of using two coupled
material sheets possessing surface mode resonances for evanescent
field enhancement is already well established in the microwave
region. This paper shows that the same principles can be applied
also in the optical region, where the performance of the resonant
sheets can be realized with the use of metallic nanoparticles. In
this paper we present design of such structures and study the
electric field distributions in the image plane of such superlens.
\end{abstract}

\maketitle

\section{Introduction}

Recently, there have been many studies of near-field enhancement
and sub-wavelength imaging using metamaterial slabs with negative
permittivity and permeability (double-negative or DNG media). The
predicted negative refraction,\cite{Veselago} which occurs at an
interface between double-positive (DPS, positive permittivity and
permeability) and DNG media, was confirmed experimentally in the
microwave domain using arrays of split rings and
wires\cite{Shelby,Parazzoli,Houck} and also using meshes of loaded
transmission lines.\cite{Eleftheriades,Caloz} Also, the predicted
enhancement of evanescent modes\cite{Pendry} was experimentally
confirmed.\cite{Grbic,Alitalo2} A lot of effort is devoted to
realization of DNG-slab superlenses in the optical
region.\cite{Zhou,Dolling,Kildishev,Zhang,Gabitov,Alu} However,
there are many obstacles on this way, due to fundamental
difficulties in realization of artificial magnetic materials in
the optical region with the use of nano-sized resonant particles.

An alternative approach to the realization of superlenses for
evanescent fields has been suggested in Ref.~16. This approach is
based on the use of a pair of coupled resonant arrays or resonant
sheets placed in a usual double-positive medium, e.g. free space
or a dielectric. Systems comprising coupled pairs of arrays of
resonant metal particles have been used to demonstrate
experimentally the sub-wavelength imaging properties at microwave
frequencies.\cite{Maslovski,Freire,Alitalo3} The main advantage of
this route to superlens design is that a superlens with
sub-wavelength resolution can be realized without using a bulk DNG
medium. Only two sheets supporting surface modes in a broad
spectrum of spatial frequencies are required, if enhancement of
only evanescent modes is desired (although the propagating modes
in this case are not focused in the image plane as with a bulk DNG
slab, the imaging is still possible
\cite{Maslovski,Freire,Alitalo3}). Removal of the bulk DNG slab
strongly mitigates the problem of losses that have been present in
any realized DNG medium so far.

The goal of the present work is to show that sub-wavelength
imaging characteristics in a device based on resonant arrays can
be achieved also at very high frequencies (the optical region and
above) if we use metallic nano-sized particles as the resonating
inclusions of the two arrays. This approach to the realization of
an optical superlens was first suggested in Ref.~19. In this paper
we will study the dispersion in two-dimensional arrays of silver
and gold nanospheres and show that the dispersion characteristics
are suitable for using these types of arrays for evanescent field
enhancement. Next, the electric field distributions in a superlens
consisting of two arrays of metal nanospheres are studied
numerically, in order to confirm and analyze the sub-wavelength
resolution of the image formed by the lens.

\section{Structure of the lens and dispersion in arrays of vertically polarized metal spheres}

\begin{figure}[h!]
\centering \epsfig{file=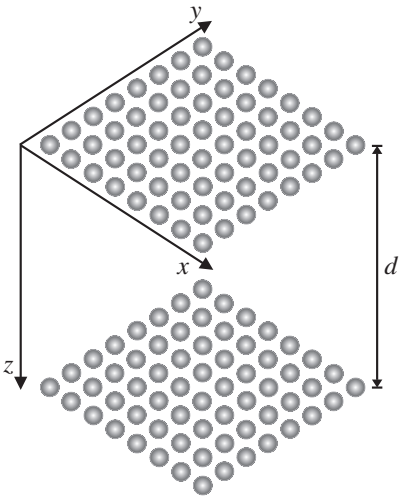, width=0.32\textwidth} \caption{A
superlens formed by two planar arrays of spherical metal
particles, embedded in a host medium.}\label{superlens}
\end{figure}

The structure of the superlens that is studied in this paper is
shown in Fig.~\ref{superlens}. The spherical particles have the diameter
which is considerably smaller than the optical wavelength, and the sphere material is a noble metal.
The spheres exhibit a plasmonic resonance
within the optical region ($\lambda_0=400$~nm...700~nm). The whole
structure (including the source and image planes) is embedded in a
host medium with the relative permittivity $\varepsilon_{\rm h}$. We
will consider the lens to be working properly if we obtain a
sub-wavelength image in the image plane with the distance between
the source and image planes being larger than $\lambda_{\rm eff}/2$
($\lambda_{\rm eff}$ is the wavelength in the host medium).

The operational principle of the device requires that at the
operational frequency each of the two sheets supports surface
waves (plasmons) in a wide range of propagation constants along
the sheet planes. Existing of eigenmodes with high values of the
propagation constant ensures resonant amplification of incident
evanescent waves with large values of the transverse wavenumber.
This means that for the optimal operation of the superlens we need
to design nano-scaled arrays whose dispersion curve is as flat as
possible in the vicinity of the operational frequency. Using the
approach of Ref.~19, we will first study the dispersion in a
single array of the proposed lens. The goal is to find the
suitable parameters of the particles and the array period using a
simplified model of infinite arrays, which can be further
optimized by numerical studies of realistic structures of a finite
size. In addition to the assumption of the infinite grid (along
$x$ and $y$), in this section we will also assume that each
particle is polarized vertically (i.e., the polarization vector is
normal to the array surface).

The dispersion in an infinite, two-dimensional array of vertical
dipoles can be calculated using the interaction coefficient $C$ of
such array:\cite{Simovski2}

\begin{center}
$C=\frac{k^2}{4\pi \varepsilon_{\rm h}
\varepsilon_0}\sum\limits_{m,n=-\infty}^{\infty}\left[1+\frac{j}{kR_{mn}}+\frac{1}{k^2R_{mn}^2}\right]$
\end{center}
\e \times\frac{e^{-jkR_{mn}}-j\textbf{k}_t\cdot
\textbf{R}_{mn}}{R_{mn}}, \l{C} \f where $k$ is the wavenumber,
$k_t=\sqrt{k_x^2+k_y^2}$ is the transverse wavenumber and $R_{mn}$
is the distance between spheres $m$ and $n$. The inverse
polarizability of a metal sphere is (e.g., Ref.~19)

\e \frac{1}{\alpha}=\frac{\varepsilon_{\rm m}+2\varepsilon_{\rm
h}}{3\varepsilon_0\varepsilon_{\rm h}\frac{4\pi
r_0^3}{3}(\varepsilon_{\rm m}-\varepsilon_{\rm h})} + j
\frac{k^3}{6\pi \varepsilon_0\varepsilon_{\rm h}}, \l{alpha1} \f
where $\varepsilon_{\rm m}$ is the permittivity of metal and $r_0$
is the radius of the sphere. The permittivity of metal can be
expressed as:\cite{Absorption}

\e \varepsilon_{\rm m}=1-\frac{\omega_p^{2}}{\omega\left(
\omega-j\omega_D \right)}, \l{perm_metal} \f where $\omega_p$ and
$\omega_D$ are the plasma and damping frequencies of the metal,
respectively. For lossless particles $\omega_D=0$ and the
dispersion equation $1/\alpha=C$ transits to the real
equation\cite{Simovski}

\e {\rm Re}\left(\frac{1}{\alpha}\right)={\rm Re}(C). \l{disp} \f

Equation \r{disp} was solved numerically using the fast-converging
representation for series \r{C}.\cite{Simovski2} By studying the
dispersion characteristics of an infinite two-dimensional array,
the dimensions of the array (i.e., the radius of the spheres and
the period $a$) can be found in such a way that the dispersion
curve is reasonably flat while the size of the spheres is of the
same order as the period of the arrays.

The parameters of the arrays that are used in this paper are shown
in Tables~\ref{table1} and~\ref{table2}, where the wavelengths
$\lambda_p$ and $\lambda_D$ correspond to the plasma and damping
frequencies of the spheres, respectively. Here we have used the
plasma and damping frequencies for bulk silver and gold, which is
an adequate approximation for the sphere sizes that we are
using.\cite{Absorption} The plasma and damping frequencies for
silver have been obtained from Ref.~22 and for gold from Ref.~21.
In the calculation of the dispersion curve, we have used
$\varepsilon_{\rm h}=1$ for simplicity. The resonant frequency
($f_{\rm r}$) of the spheres that is shown in Tables~\ref{table1}
and~\ref{table2} is calculated from the plasma frequency with

\e \omega_r=\omega_p/\sqrt{1+2\varepsilon_{\rm h}}. \f

\begin{table}[h]
\centering \caption{Parameters of the silver sphere array.}
\label{table1}
\begin{tabular}{|c|c|c|c|c|} \hline
     $r_0$   &  $a$  &  $\lambda_p$  & $\lambda_D$ & $f_{\rm r}$ ($\varepsilon_{\rm h}=1$)\\

\hline

28 nm & 65 nm & 328 nm & 58433 nm & $5.2806\cdot 10^{14}$ Hz \\

\hline
\end{tabular}
\end{table}

\begin{table}[h]
\centering \caption{Parameters of the gold sphere array.}
\label{table2}
\begin{tabular}{|c|c|c|c|c|} \hline
     $r_0$   &  $a$  &  $\lambda_p$  & $\lambda_D$ & $f_{\rm r}$ ($\varepsilon_{\rm h}=1$)\\

\hline

15 nm & 40 nm & 145 nm & 11500 nm & $1.1945\cdot 10^{15}$ Hz \\

\hline
\end{tabular}
\end{table}

\begin{figure}[h!]
\centering \epsfig{file=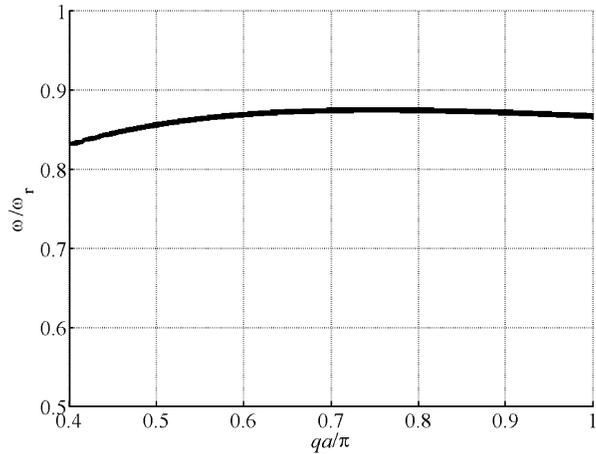, width=0.55\textwidth}
\caption{Dispersion of an infinite two-dimensional array of silver
nanospheres. Normalized frequency as a function of the normalized
transverse wavenumber. The spheres are assumed to be polarized
vertically.}\label{dispersion_silver}
\end{figure}

\begin{figure}[h!]
\centering \epsfig{file=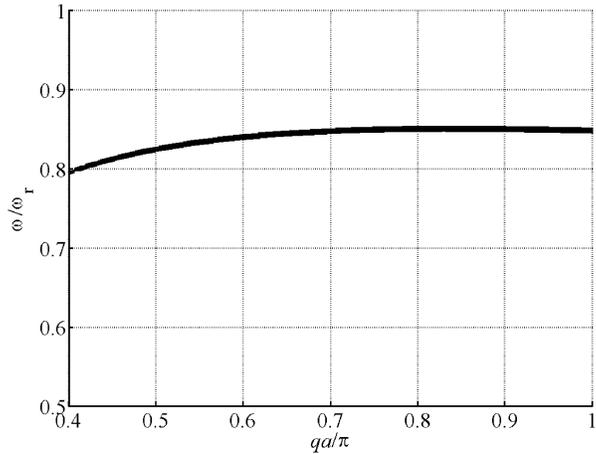, width=0.55\textwidth}
\caption{Dispersion of an infinite two-dimensional array of gold
nanospheres. Normalized frequency as a function of the normalized
transverse wavenumber. The spheres are assumed to be polarized
vertically.}\label{dispersion_gold}
\end{figure}

With the values shown in Tables~\ref{table1} and~\ref{table2}, the
dispersion curves of the arrays can be plotted using \r{disp}, see
Figs.~\ref{dispersion_silver} and~\ref{dispersion_gold}. From
these results we can conclude that the dispersion curve for both
types of metal spheres is reasonably flat in a large range of
values of $q$ (where $q=k_x/\sqrt{2}=k_y/\sqrt{2}$).

\section{Field distributions in the image plane of the lens}

\begin{figure}[h!]
\centering \epsfig{file=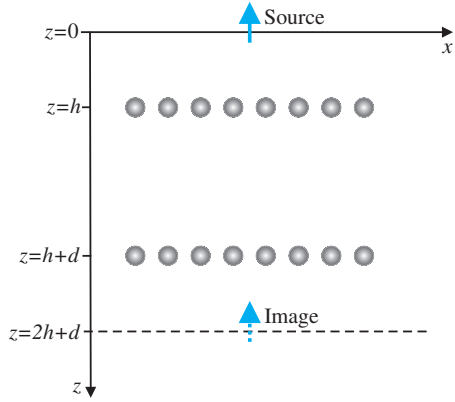, width=0.36\textwidth}
\caption{(Color online) Ideal operation of the
superlens.}\label{superlens2}
\end{figure}

With the suitable array dimensions found in the previous section,
it is now possible to study the electric field distributions in a system of
two \textit{finite} two-dimensional arrays of resonant
nanospheres. The losses of the metal spheres are taken into account
by using complex values for the polarizability of the spheres and the permittivity of the metal.
The ideal operation of the lens is illustrated in
Fig.~\ref{superlens2}, where the image (defined in the image
plane, i.e., $z=2h+d$) appears as a perfect reconstruction of the
source (situated in the plane $z=0$).

The fields are calculated simply by considering each sphere as
having three orthogonal dipole moments at the same frequency
(where the dispersion curve is flat) and calculating separately
the contribution of each sphere to the image plane field. Because
the two arrays of spheres interact, the first task is to solve all
the dipole moments $\textbf{P}_m$ that correspond to each
sphere, taking into account the interaction between all the
spheres. This can be done by solving the following equation:

\e \textbf{P}_m = \alpha
\textbf{E}_{\rm loc}=\alpha\left[\textbf{E}_m^{\rm ext} + \sum_{n\neq
m}\textbf{E}_n \right], \l{Pm} \f
where $\textbf{E}_{\rm loc}$ is the
local field acting on the $m$:th sphere, $\textbf{E}_m^{\rm ext}$ is
the external field (caused by the source) and
$\textbf{E}_n={\overline{\overline\Phi}}_{m,n}\cdot\textbf{P}_n$ is the
field caused by dipole $\textbf{P}_n$. All fields are evaluated at
the position of the $m$:th sphere.
Here the dyadic function ${\overline{\overline\Phi}}_{m,n}$ (with $m\neq n$)
describes the interaction between spheres $m$ and $n$. If we
introduce the notation \e
{\overline{\overline\Phi}}_{m,m}=-\frac{1}{\alpha}\overline{\overline{I}},
\f where $\overline{\overline{I}}$ is the unit dyadic, \r{Pm} can
be expressed in a simpler form (for each orthogonal component of
the vectors separately):

\e \sum_{n}\Phi_{m,n}P_n=-E_m^{\rm ext}. \l{sum_n} \f The electric
field at point ($x$, $y$, $z$) of a dipole with the dipole moment
$\textbf{p}$ placed at ($x'$, $y'$, $z'$) is (e.g., Ref.~23)

\begin{center}
$\textbf{E}=\frac{1}{4\pi \varepsilon_{\rm h}\varepsilon_0}\left\{
k^2(\textbf{u}\times \textbf{p})\times
\textbf{u}\frac{e^{-jkr}}{r}\right\}$\\ \end{center} \e
+\frac{1}{4\pi\varepsilon_{\rm h}\varepsilon_0}\left\{[3\textbf{u}(\textbf{u}\cdot\textbf{p})-\textbf{p}]\left(\frac{1}{r^3}+\frac{jk}{r^2}\right)e^{-jkr}\right\},
\l{E} \f where

\e
\textbf{u}=\frac{(x-x')\textbf{x}_0+(y-y')\textbf{y}_0+(z-z')\textbf{z}_0}{r}
\f and

\e r=\sqrt{(x-x')^2+(y-y')^2+(z-z')^2}. \f
From \r{E} we can
derive ${\overline{\overline\Phi}}_{m,n}$.

Next, let us assume that two finite two-dimensional arrays of
spheres are placed in a host medium with permittivity
$\varepsilon_{\rm h}$. The distance between the arrays along the
$z$-direction is $d$. Also, let us assume that a source, which
consists of one or more short vertical electric dipoles, is placed
on top of the first array at  distance $h$ from the surface of
that array. The source excites both arrays, and the $x$, $y$ and
$z$-components of the field at the position of each sphere
($E_{m,x}^{\rm ext}$, $E_{m,y}^{\rm ext}$ and $E_{m,z}^{\rm ext}$)
can be calculated from \r{E} by choosing
$\textbf{p}=p\textbf{z}_0$. The $x$-, $y$-, and $z$-components of
$\Phi_{m,n}$ can also be calculated with \r{E}, where $r$ is now
the distance from sphere $m$ to $n$. The dipole moments $P_{n,x}$,
$P_{n,y}$ and $P_{n,z}$ of each sphere can then be calculated from
\r{sum_n}. With the dipole moments solved, \r{E} can be used to
calculate the vertical component of electric field in the image
plane caused by all the spheres in the two arrays. To get the
total field in the image plane, we must add also the field
produced by the source to the field of the arrays.

We have calculated the field distributions in the image plane for
different sphere materials (silver and gold) and also for
different sources (one or more vertical dipoles in the source
plane) using \r{sum_n} and \r{E}. In the following, we will study
arrays with 20 $\times$ 20 silver spheres in each array. The
dimensions of the arrays are the same as shown in
Table~\ref{table1}, i.e., the radius of the spheres is $r_0=28$~nm
and the period of the arrays is $a=65$~nm. The dimensions of the
lens are $h=a$ and $d=2a$. If we choose the source plane to be at
$z=0$, then we plot the image plane field at $z=d+2h=4a$. As the
permittivity of the host material we have used $\varepsilon_{\rm
h}=2.301$, which corresponds to the permittivity of PMMA
(polymethyl methacrylate) dielectric.\cite{Lee} PMMA is used here
due to its very low losses (in fact, we have neglected the losses
in the host material to simplify the calculations).

Without any extensive optimization procedure, we have found a
suitable frequency of operation to be $1.037f_{\rm r}$ (where
$f_{\rm r}\approx 3.8643\cdot 10^{14}$ Hz), which corresponds to
the effective wavelength of $\lambda_{\rm eff}\approx 493.52$~nm
in the host medium. Comparison of this result with the dispersion
curve in Fig.~\ref{dispersion_silver} shows a considerable
difference in the expected operational frequency. This effect can
be explained by the fact that the curve in
Fig.~\ref{dispersion_silver} is plotted for an \textit{infinite}
array. Indeed, as the number of the spheres in the array
increases, the operational frequency is expected to decrease, as
in arrays of resonant scatterers.\cite{Tretyakovbook} At the
frequency $1.037f_{\rm r}$, the distance between the source and
image planes is $4a=260$~nm $\approx0.53\lambda_{\rm eff}$.

\subsection{Excitation by a single source}

\begin{figure}[h!]
\centering \epsfig{file=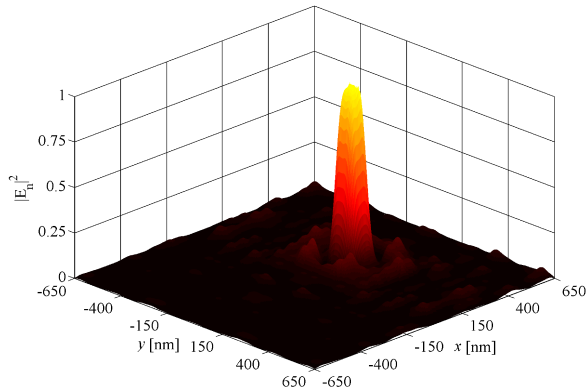, width=0.5\textwidth}
\caption{(Color online) Calculated electric field distribution in
the image plane with a single source. Fields are normalized to the
maximum value.}\label{1source3d}
\end{figure}

\begin{figure}[h!]
\centering \epsfig{file=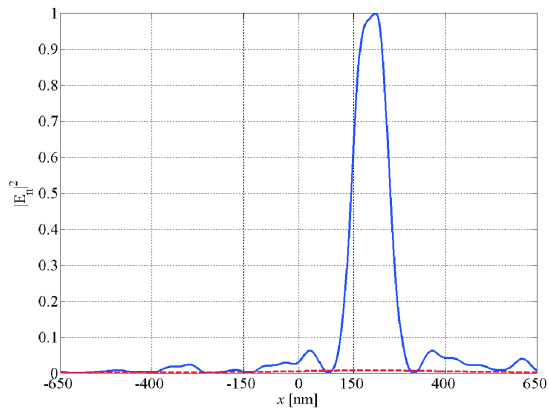, width=0.5\textwidth}
\caption{(Color online) Calculated electric field distributions
along the line $y=0$~nm in the image plane. Fields are normalized
to the maximum value. Solid line: image plane field with the
arrays. Dashed line: image plane field without the
arrays.}\label{1source2d}
\end{figure}

First, let us have a look at the electric field distribution in
the image plane caused by a point source (a short vertical electric dipole)
which is situated in the source plane. In this example, the position of the source dipole is at $x=3a$,
$y=0$ (the origin of the coordinate system is now at the center of
the arrays in the $xy$-plane). See Fig.~\ref{1source3d} for the
distribution of the $z$-component of the electric field plotted in
the image plane, i.e., in the plane $z=2h+d=4a$. For a more
detailed picture of the field distribution, see
Fig.~\ref{1source2d}, where we have plotted the field
distributions along the line $y=0$.

From Fig.~\ref{1source2d} we can conclude that the half-power
width of the ``image'' is about $0.23\lambda_{\rm eff}$, which clearly
confirms the sub-wavelength imaging effect. As can also be seen from
Fig.~\ref{1source2d}, the field in the image plane is very strongly
enhanced by the arrays: The field in the image plane without the
arrays (dashed line) is negligible compared to the field strength of the
image formed by the ``lens''.

\subsection{Excitation by two sources}

\begin{figure}[h!]
\centering \epsfig{file=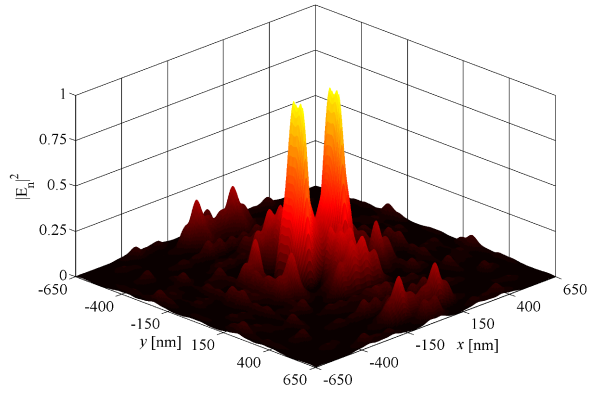, width=0.5\textwidth}
\caption{(Color online) Calculated electric field distribution in
the image plane with two sources. Fields are normalized to the
maximum value.}\label{2sources3d}
\end{figure}

\begin{figure}[h!]
\centering \epsfig{file=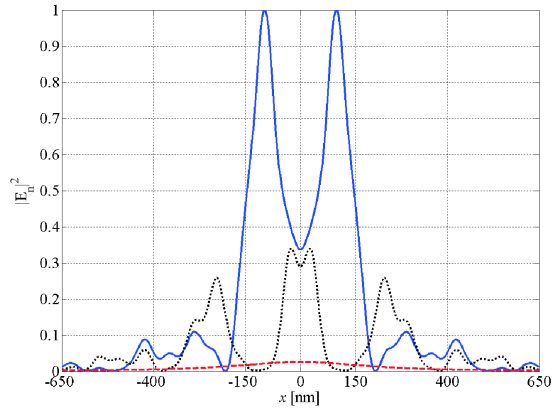, width=0.5\textwidth}
\caption{(Color online) Calculated electric field distributions in
the image plane. Fields are normalized to the maximum value. Solid
line: $y=0$~nm, image plane field with the arrays. Dashed line:
$y=0$~nm, image plane field without the arrays. Dotted line:
$y=106$~nm, image plane field with the arrays.}\label{2sources2d}
\end{figure}

Next, let us consider excitation by two short vertical dipoles,
situated in the source plane. To study the resolution properties
of the lens, we place the sources very close to each other. The
position of source dipole (1) is $x=1.25a=81.25$~nm, $y=0$~nm and
the position of source dipole (2) is $x=-1.25a=-81.25$~nm, $y=0$
nm. With this positioning the distance between sources (1) and (2)
is approximately $0.329\lambda_{\rm eff}$. See
Fig.~\ref{2sources3d} for the distribution of the $z$-component of
the electric field plotted in the image plane. For a more
qualitative picture of the field distribution, see
Fig.~\ref{2sources2d}, where we have plotted the field
distribution along the line $y=0$.

From Fig.~\ref{2sources2d} we can conclude that the two sources
can be resolved very reliably (on the level half of the maximum
intensity) from the image plane field distribution. As can be seen
from Fig.~\ref{2sources3d}, the introduction of multiple sources
causes some additional maxima (because of the interference
effect), which can potentially hinder the resolution in the image
plane. The maximum field corresponding to these unwanted maxima
(along line $y=106$~nm) is also plotted in Fig.~\ref{2sources2d}
(dotted line). We see that this maximum is well below the
half-power level of the total intensity.

\subsection{Effect of the positioning of the sources}

The non-symmetric positioning of the sources with respect to the
unit cell of the arrays affects the field distributions. The
cases studied above relate to the ``worst case'', because there
the sources are positioned in the center between four neighboring
spheres of the arrays (in the $xy$-plane). When a source is
positioned directly above a sphere, the maximum of the image will
be even more pronounced. Also, when using two sources that are not
symmetrically positioned with respect to the unit cell of the
arrays (as in the previous subsection), it may happen that the
image of the other source has a larger maximum (which corresponds
to the fact that this source is closer to a sphere in the
$xy$-plane). By studying these special cases it was noticed that
the effect of the source positioning is not crucial to the
formation of a clear and unambiguous image. The maxima
corresponding to the positions of the sources are always above the
half-power level.

The effect of the positioning of the sources with respect to the
entire sphere array was also studied. It has been noticed that even
when using the arrays of 20 $\times$ 20 spheres, the image is
properly formed as long as the sources are not very close to the
edge of the arrays (in the $xy$-plane). When using two or more
sources, the interference maxima grow stronger as the sources get
closer to the edge of the arrays.

\begin{figure}[h!]
\centering \epsfig{file=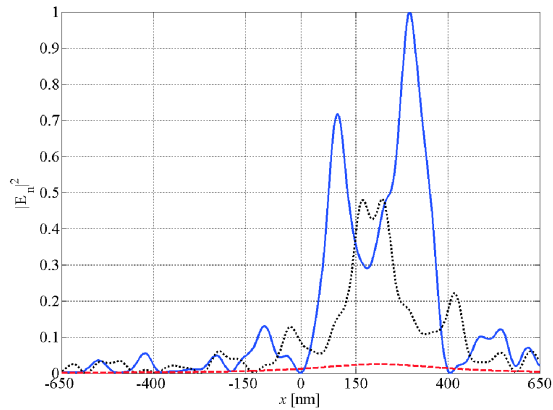, width=0.5\textwidth}
\caption{(Color online) Calculated electric field distributions in
the image plane. Fields are normalized to the maximum value. Solid
line: $y=325$~nm, image plane field with the arrays. Dashed line:
$y=325$~nm, image plane field without the arrays. Dotted line:
$y=167$~nm, image plane field with the
arrays.}\label{2sources2d_v2}
\end{figure}

For an example, see Fig.~\ref{2sources2d_v2}, where two sources
are positioned in such a way that the distance from the sources to
their closest spheres is different and the both sources are close
to the edge of the arrays (in the $xy$-plane). The position of
source dipole (1) is $x=4.5a=292.5$~nm, $y=5a=325$~nm and the
position of source dipole (2) is $x=2a=130$~nm, $y=5a=325$~nm.
With this positioning the distance between the sources is the same
as in the above example (approximately $0.329\lambda_{\rm eff}$).

The reason for the difference in the amplitudes of the two
``images'' in Fig.~\ref{2sources2d_v2} is not the fact that one
source is closer to the edge of the array than the other. The
main reason for this is the positioning of the sources with respect to
the unit cell of the arrays. However, the interference maximum is
affected also by the distance to the edge of the arrays. The dotted
line in Fig.~\ref{2sources2d_v2} corresponds to this interference term of
the total field. The maximum of this interference term is somewhat
stronger than the one in Fig.~\ref{2sources2d}, and it is due to
the placing of the sources near the edge of the arrays.

\subsection{Further improvements of the resolution of the proposed
lens}

It is possible to further improve the resolution characteristics
of the lens studied in this paper. First, the removal of the
propagating modes from the lens should mitigate the unwanted
interference maxima (this can be realized simply by introducing a
thin silver slab between the source plane and the lens). Second
way to improve the imaging properties is to introduce a small
deviation in the parameters of the arrays (i.e., the radius and
the period). This will make the dispersion curve even more flat, with
the result that more modes are supported by the arrays. The study
of these improvements is beyond the scope of this paper.

\section{Conclusions}

In this paper we have studied the possibility of using a coupled
pair of arrays comprising metallic nanospheres as a near-field
imaging device enhancing evanescent fields. We have shown that in
arrays with infinitely many silver or gold spheres the dispersion
is flat enough so that in a very narrow frequency band most of the
evanescent modes are resonantly excited in the arrays. According
to the studies made in this paper, this excitation enables the
``superlensing'' effect, already known in the microwave region, in
the optical domain. In the proposed device the enhancement of a
large number of the evanescent modes emitted by the source can be
realized. We have numerically studied a superlens consisting of
two finite arrays of silver spheres and have shown that in the
image plane of the lens, resolution better than $\lambda_{\rm
eff}/3$ is achievable, even when the distance between the source
and image planes is larger than $\lambda_{\rm eff}/2$. According
to the results presented in this paper, the use of metallic
nanospheres is a very prospective way of extending the use of
near-field enhancement phenomenon into the optical region.

\section*{Acknowledgments}

This work has been partially funded by the Academy of Finland and
TEKES through the Center-of-Excellence program. The authors wish
to thank Liisi Jylh$\ddot{\rm a}$ for helpful discussions.

\end{document}